\documentclass{ws-ijmpe}

\newcommand{\eq}{\begin{eqnarray}}
\newcommand{\en}{\end{eqnarray}}

\newcommand{\delslash}{\partial \!\!\! /}  
\def \etal{{\it et al.\,\,}}
\def \ie{{\it i.e.\,\,}}
\def \mate<#1|#2|#3>{\mbox{$\langle {#1}|\,{#2}\,|{#3}\rangle$}}

\begin{document}

\title{Mass spectrum of the $J^P=1/2^-$ and $3/2^-$ pentaquark antidecuplets
in the perturbative chiral quark model}
                
\author{T. Inoue, V. E. Lyubovitskij, Th. Gutsche and Amand Faessler}

\address{Institut f\"ur Theoretische Physik, Universit\"at T\"ubingen,\ 
Auf der Morgenstelle 14,  D-72076 T\"ubingen, Germany }

\maketitle

\begin{history}
\received{(received date)}
\revised{(revised date)}
%\accepted{(Day Month Year)}
%\comby{(xxxxxxxxxx)}
\end{history}

\begin{abstract}
We study the recently discovered $\Theta^+$ baryon in the
context of the perturbative chiral quark model.
The basic configuration of the $\Theta^+$ is set up as a
pentaquark bound state, where the single particle wave
functions are the ground state solutions of a confining potential.
We classify the resulting pentaquark multiplets as the $J^P=1/2^-$ and
$3/2^-$ flavor SU(3) antidecuplet. The full mass spectrum of the multiplets 
is determined by including meson and gluon cloud contributions inducing
flavor SU(3) breaking.
Mainly due to the semi-perturbative
gluon effects
the resulting $3/2^-$ antidecuplet is about 185 MeV heavier
than the $1/2^-$ one.
We assign the observed $\Theta^+$ baryon as a member of 
the $1/2^-$ antidecuplet and discuss in particular the relation
to the recent experimental signal for a $\Xi^{--}$ baryon.
\end{abstract}

%\pacs{12.39Fe, 12.39.Ki, 12.40.Yx, 14.20.-c} 
%\keywords{$\Theta^+$ baryon; Pentaquark state; Antidecuplet; 
%Relativistic quark model; Chiral Lagrangians}

\section{Introduction}\label{Sect_Intro} 

Recently an exotic baryon, which is now called the $\Theta^+$, was observed 
by several experimental collaborations~\cite{Nakano:2003qx,Stepanyan:2003qr}.  
This baryon is supposed to be an exotic hadron with positive 
strangeness $S=+1$. Since such a quantum number requires the additional
existence of at least a valence antiquark in the $\Theta^+$ baryon,
this minimal configuration is in clear contrast to 
the three valence quark picture of conventional baryons. 
Another characteristic property of the $\Theta^+$ is its unusually narrow 
decay width. From recent experiments upper bounds for the total width of
the $\Theta^+$ were reported to be around $9 \sim 25$ MeV. Moreover, a recent 
revised analysis of kaon-deuteron scattering data~\cite{Arndt:2003dm}  
shows that the $\Theta^+$ width should be smaller than 1 MeV if 
the baryon exists and if the isospin assignment equals zero.
There is also an experimental (very preliminary) indication that
the $\Theta^+$ is a isosinglet~\cite{Stepanyan:2003qr}. 
After the observation of the $\Theta^+$ another narrow exotic baryon,
the $\Xi^{--}$, with double strangeness $S=-2$ was recently claimed
to be seen~\cite{Alt:2003vb}. 
  
Many theoretical interpretations have been put forward
for the $\Theta^+$ baryon. 
Originally, the $\Theta^+$ baryon was predicted by the chiral quark soliton
model~\cite{Diakonov:1997mm} including its mass and narrow decay 
width, before it was confirmed by the 
experiments~\cite{Nakano:2003qx,Stepanyan:2003qr}. 
In this model baryons are considered as bound states of valence quarks 
in a self-consistent field of pseudoscalar mesons.  
The $\Theta^+$ baryon shows up as a isosinglet member of the spin-parity 
multiplet $J^P=1/2^+$ which is classified as a flavor SU(3) antidecuplet 
$\bar{10}_f$. Note that the prediction of the $\Theta^+$ baryon which 
is based on a collective quantization of the chiral soliton was shown to be 
inconsistent with large N$_c$ QCD counting~\cite{Cohen:2003yi}. 

The interpretation of the $\Theta^+$ as a quasi-bound $\pi K N$
system with isospin 
$I=0$ and spin-parity $J^P = 1/2^+$ was tested in 
Ref.~\cite{Llanes-Estrada:2003us}. 
The analysis in Ref.~\cite{Llanes-Estrada:2003us} 
is based on a solution of a partial Faddeev equation in the chiral 
unitary approach by assigning the $\kappa$ resonance to the $K \pi$ 
subsystem. The authors did not obtain a bound state and concluded 
that the $\Theta^+$ baryon is unlikely to be a quasi-bound $\pi K N$ system.

Another popular scenario for the structure of the $\Theta^+$ is, of
course, a pentaquark picture, where the minimal configuration of
four quarks and a single antiquark form a bound state. Many models for  
the $\Theta^+$ baryon as a pentaquark state have been developed since
the first evidence for this state was reported.
They can be divided into two main categories, namely
correlated~\cite{Jaffe:2003sg,Karliner:2003sy} and 
uncorrelated quark models~\cite{Capstick:2003iq,Stancu:2003if,Carlson:2003pn,Carlson:2003wc,Jennings:2003wz}.
In the first kind of approaches strongly correlated subsystems of two or
three quarks (and the antiquark)
are treated as single particles. Such an ansatz leads to a considerable
simplification of the inner structure of exotic baryons in terms of
three- or two-body systems which is furthermore exploited.

A typical example for a correlated model is the one originally
proposed by Jaffe and Wilczek~\cite{Jaffe:2003sg}.
They suggested that the $\Theta^+$ 
baryon is a bound state of two diquarks and a single antiquark,  
where each diquark has spin-parity $0^+$ in 
a $\bar 3_c$ color and a $\bar 3_f$ flavor state. These diquarks 
obey Bose statistics and their spatially antisymmetric wave function 
contains a P-wave in angular momentum. Exotic baryons (including the 
$\Theta^+$) result from a bound system of two diquarks and 
a single antiquark with spin-parity either $1/2^+$ or $3/2^+$. 
Choosing the total spin-parity as 
$1/2^+$ and ideal mixing for the resulting $8_f$ and $\bar{10}_f$ flavor 
multiplets, the lightest member of the exotic multiplet is identified
with the Roper resonance with a fixed mass of 1440 MeV. The parity 
of the $\Theta^+$ is positive since one unit of orbital 
angular momentum is involved, which is compensated
by the negative intrinsic parity of the antiquark.

Karliner and Lipkin considered an alternative scheme of a 
correlated quark model. Their approach is based on the idea that
the $\Theta^+$ baryon is considered as a bound 
state of a di- and a
triquark~\cite{Karliner:2003sy} coupled together in a relative $P$-wave. 
The triquark is proposed to be the correlated system of two quarks 
and a single antiquark with total spin-parity $1/2^-$
in the $3_c$, $\bar 6_f$ state.  
The spin-parity of the total correlated five-quark system is 
$1/2^+$ and $3/2^+$ as in the 
Jaffe-Wilczek model. Moreover, the $\Theta^+$ is again 
a member of the $\bar{10}_f$ multiplet arising from a $8_f + \bar{10}_f$ 
decomposition. The mass of the $\Theta^+$ is estimated 
phenomenologically by using input from the $D$ meson spectrum. 

In uncorrelated quark models the five-body system is studied directly,
mostly in a direct extension of the conventional baryon system consisting
of three valence quarks. 
Several models based on different valence configurations and
varieties in the residual dynamics have been proposed and 
studied~\cite{Capstick:2003iq,Stancu:2003if,Carlson:2003pn,Carlson:2003wc,Jennings:2003wz}.
This traditional scenario for the pentaquark structure of the $\Theta^+$ 
is claimed by several authors involved in
correlated models to be problematic. In fact, because of unconstraint number
of degrees of freedom the uncorrelated five-body approaches 
lead to a larger number of possible configurations of constituents 
than correlated ones.
On the other hand, uncorrelated models cover a wide spectrum 
of possibilities for the possible pentaquark structure of the
$\Theta^+$ baryon.
For example, it is possible to construct a negative parity antidecuplet 
within uncorrelated models, while it is difficult to be realized 
using the correlated approaches discussed above. The parity of the 
$\Theta^+$ baryon is not experimentally determined yet. If the
parity is negative, the models suggesting a positive parity for the  
$\Theta^+$ baryon are ruled out immediately. 
In particular, QCD sum rule calculations~\cite{Zhu:2003ba,Sugiyama:2003zk}
and a lattice QCD simulation~\cite{Sasaki:2003gi} 
indicate a negative parity for the $\Theta^+$.
The main advantage of the five-body treatment is that 
the particles involved are ordinary ones, \ie quarks and antiquarks, 
which are the same degrees of freedoms as in conventional hadrons.
A straightforward extrapolation of the conventional hadron systems
to their exotic partners helps to obtain a unified understanding
of the complete baryon spectrum.
It is relatively easy to introduce and define the dynamics in such models 
by assuming a universal underlying dynamical picture
in explaining the full set of baryon states.
Moreover, in this treatment the quark Fermi statistics can be 
imposed strictly, while in the correlated approaches
it is only exactly fulfilled
when the diquark is really a pointlike particle. 

In this paper, we construct an uncorrelated pentaquark picture for the 
$\Theta^+$ baryon in extension of the perturbative chiral quark 
model~\cite{Lyubovitskij:2001nm}. This quark model was originally developed
as an effective approach to 
conventional baryons considered as bound states of three 
valence quarks which are supplemented by a cloud of pseudoscalar mesons. 
The model has been successfully applied to the description of canonical baryon
properties~\cite{Lyubovitskij:2001nm,Lyubovitskij:2000sf,Inoue:2004}.
In the current work, we tune the model such that it can be extended
to the pentaquark 
systems. We then investigate the mass spectrum of the negative
parity antidecuplets with spin 1/2 and 3/2, where one of the members
can be associated with the $\Theta^+$ baryon.

\section{The perturbative chiral quark model (PCQM)}
\label{Sect_PCQM} 

The perturbative chiral quark 
model~\cite{Lyubovitskij:2001nm,Lyubovitskij:2000sf,Inoue:2004} 
is aimed at the description of baryons based on an effective chiral
Lagrangian.
The model describes the valence quarks of baryons as relativistic fermions 
moving in an external field (static potential) 
$V_{\rm eff}(r)=S(r)+\gamma^0 V(r)$ with $r=|\vec x|$.   
The valence quark core is supplemented in the flavor SU(3) version
by a cloud of Goldstone 
bosons $(\pi, K, \eta)$ according to the 
chiral symmetry requirement and in addition by quantum fluctuations of 
the gluon field~\cite{Leutwyler:1980ma}. Treating also Goldstone fields as 
small fluctuations around the valence quark core, we derive the 
linearized effective chiral Lagrangian~\cite{Inoue:2004}: 
\begin{eqnarray}\label{linearized_L}
{\cal L}_{\rm eff}(x) 
 &=&   \bar{\psi}(x) \left[ i\,\delslash - V_{\rm eff}(r) \right]\psi(x)
     + \frac{1}{2} \sum\limits_{i=1}^{8} [\partial_\mu \Phi_i(x)]^2 
     - \frac{1}{4} F^a_{\mu\nu} F^{a \, {\mu\nu}} \nonumber\\ 
 &-& \bar{\psi}(x) \biggl\{ S(r) i \gamma^5 \frac{\hat \Phi (x)}{F} + 
     g_s \gamma^\mu A_\mu^a(x) \frac{\lambda^a}{2} \biggr\}\psi(x)
     + {\cal L}_{\chi SB}(x)
\end{eqnarray} 
where $F=88$ MeV is the pion decay constant in the chiral
limit\cite{Gasser:1987rb}; $g_s$ is the quark-gluon coupling constant; 
$A_\mu^a$ is the quantum component of the gluon field and $F^a_{\mu\nu}$ is 
its conventional field strength tensor;   
$\hat\Phi = \sum\limits_{i=1}^{8} \Phi_i \lambda_i = \sum\limits_P 
\Phi_P \lambda_P$ is the octet matrix of pseudo-scalar mesons with 
$P = \pi^\pm, \pi^0, K^\pm, K^0, \bar K^0, \eta$. 
The term ${\cal L}_{\chi SB}(x)$ in Eq.~(\ref{linearized_L}) contains the mass 
contributions both for quarks and mesons, 
which explicitly break chiral symmetry, 
\begin{equation}
{\cal L}_{\chi SB}(x) = - \bar\psi(x) {\cal M} \psi(x)
- \frac{B}{2} Tr [\hat \Phi^2(x)  {\cal M} ]\,.
\end{equation} 
Here, ${\cal M}={\rm diag}\{m_u,m_d,m_s\}$
is the mass matrix of current quarks, 
$B=-\langle 0|\bar u u|0 \rangle / F^2$ 
is the quark condensate constant.
We rely on 
the standard picture of chiral symmetry breaking~\cite{Gasser:1982ap}
and use the leading term for the masses of the pseudoscalar mesons. 
In our numerical calculations, we restrict to the isospin symmetry 
limit with $m_u=m_d=\hat m$. After diagonalization the physical meson masses
are given by 
\begin{equation}\label{M_Masses}
M_{\pi}^2=2 \hat m B, \hspace*{.5cm} M_{K}^2=(\hat m + m_s) B,
\hspace*{.5cm} M_{\eta}^2= \frac{2}{3} (\hat m + 2m_s) B\,.
\end{equation}
In our evaluation, we choose the following set of QCD 
parameters~\cite{Gasser:1982ap}:
\begin{equation}
\hat m = 7 \;{\rm MeV},\; \frac{m_s}{\hat m}=25,\;
B = \frac{M^2_{\pi^+}}{2 \hat m}=1.4 \;{\rm GeV}.
\end{equation}
We formulate perturbation theory in the expansion parameter 
$\hat\Phi(x)/F \sim 1/\sqrt{N_c}$ and treat finite current quark 
masses perturbatively~\cite{Lyubovitskij:2001nm}. 
All calculations are performed at one loop or at order of accuracy 
$o(1/F^2, \hat{m}, m_s)$. 

The unperturbed eigenstates of quarks and antiquarks in the confining 
potential are determined by the Dirac equation 
\begin{equation}
\left[i \delslash - V_{\rm eff}(r) \right]\psi(x) = 0 \,.
\label{eqn:diraceq}
\end{equation}
We denote the quark and the antiquark wave functions in the
respective orbits $\alpha$ and $\beta$ as:
\begin{equation}
 u_{\alpha}(x)=u_{\alpha}(\vec x)\exp(-i{\cal E}_{\alpha}) \, , ~~
 v_{\beta}(x)=v_{\beta}(\vec x)\exp(i{\cal E}_{\beta}) \,.
\end{equation}
The eigenstates are used for the expansion of the quark field
$\psi(x)$ as
\begin{equation}\label{total_psi}
 \psi(x) = \sum\limits_\alpha b_\alpha  
u_\alpha(\vec{x})\exp(-i{\cal E}_\alpha t)
         + \sum\limits_\beta  d_\beta^{\dagger} 
v_\beta(\vec{x}) \exp(i{\cal E}_\beta t)
\end{equation}
where $b_\alpha$ and $d_\beta^{\dagger}$ are the corresponding
single quark annihilation and antiquark creation operators.

We use the single-particle ground state configurations,
namely $u_0(x)$ and $v_0(x)$,
to construct the unperturbed baryonic state $|\phi_0 \rangle^B$. 
For example, the wave function of conventional baryons is given by
the direct product of single quark wave functions, \ie by  
$\prod\limits_{i=1}^{3}u_0(x_i)$ in analogy to the non-relativistic 
non-interacting quark model.
The unperturbed wave function of the $\Theta^+$ 
baryon is then set up as $\prod\limits_{i=1}^{4}u_0(x_i) \bar v_0(x_5)$ 
in our pentaquark picture. Here, color, spin and flavor indices are 
suppressed. The explicit form of the ground state quark wave function is
set up as
\begin{equation}\label{uv_0}
 u_0(\vec r) = \left( \begin{array}{c} 
                           g(r)\\
                        -i f(r) \vec\sigma\cdot\hat r
                      \end{array}
               \right) Y^0_0(\hat r) \chi_s \chi_f \chi_c, 
\label{eqn:quarkwave}
\end{equation}
while for the ground state of the antiquark we have
\begin{equation}
 v_0(\vec r) = \left( \begin{array}{c} 
                         - l(r) \vec\sigma\cdot\hat r \\
                         i k(r)
                      \end{array}
               \right) Y^0_0(\hat r) \chi_s \chi_f \chi_c ~,
\label{eqn:aquarkwave}
\end{equation}
where in $v_0(\vec r)$ the dominant lower component is an S-wave.
The radial wave functions $f(r)$, $g(r)$, $k(r)$ and $l(r)$ 
are determined by the Dirac equation (\ref{eqn:diraceq}) and depend 
on the choice of the confining potential.
 
\begin{figure}[tb]
\begin{center}
\includegraphics[width=8cm]{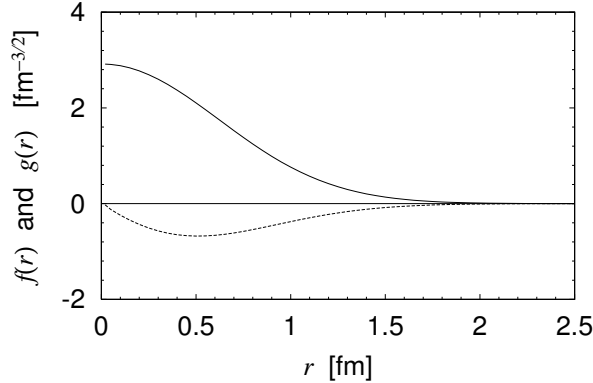}
\end{center}
\caption{\label{fig:positive}Radial wave function for the valence quark:
          solid line for upper $g(r)$ and dashed line for lower component $f(r)$ .}
\end{figure}

Previously we have used a three-parameter harmonic oscillator
potential to set up the single-particle ground state
wave functions~\cite{Inoue:2004}.
In the present approach we generalize the confining potential to also
include antiquark solutions.
First we use a linear type scalar potential 
$S(r) = c \, r $ with $c = 0.11$ GeV$^2$.
Here, the strength constant $c$ is chosen such that the resulting quark wave
functions yield a value for the charge radius of the proton with
$\sqrt{<r_E^2>^P_{LO}} = 0.76$ fm at tree level as for the harmonic
case~\cite{Inoue:2004}.
We furthermore introduce for simplicity a vector potential
with $V(r) = V_0 = {\rm constant}$.
A constant vector potential merely shifts the energy of the eigenstates
which only appears in the absolute values of the resulting baryon masses.
For $V(r) = V_0$ the quark and antiquark wave 
functions are symmetric with 
\begin{eqnarray}\label{choice_kl}  
g(r)=-k(r)  \ \ {\rm and}  \ \ f(r)=-l(r)\,. 
\end{eqnarray}  
Here the minus signs just occur due to our convention in defining
the radial wave functions of Eqs.~(\ref{eqn:quarkwave}) 
and (\ref{eqn:aquarkwave}): $g(r)$ and $l(r)$ are chosen to be positive.
The radial wave functions for the quark are shown in Fig.~\ref{fig:positive}. 
The single energies of quark and antiquark are given 
by $\pm 537 \,\mbox{MeV} + V_0$, respectively. Previously~\cite{Inoue:2004} 
we also introduced a slight flavor dependence in the effective potential 
resulting in an improved fit of hyperon properties
\ie masses and weak decay parameters. 
In this paper, for simplicity,  we do not introduce such a difference,
hence we refer to the ``symmetric parameter'' version of~\cite{Inoue:2004}.

The expectation value of an operator $\hat A$ in baryon B is set up as:
\begin{equation}\label{perturb_A}
<\hat A> = {}^B\!\langle \phi_0 |
    \sum^{\infty}_{n=0} \frac{i^n}{n!}\int \! d^4 x_1 \ldots d^4 x_n 
    T[{\cal L}_I (x_1) \ldots {\cal L}_I (x_n) \hat A\,]|\phi_0 
\rangle^B_c 
\end{equation}
where ${\cal L}_I (x)$ refers to the interaction Lagrangian,
\ie the 4-th term of the total Lagrangian of Eq.~(\ref{linearized_L}),
and subscript c refers to connected graphs only.
The interaction Lagrangian includes effects of the meson cloud and
gluonic quantum corrections to the baryon.
For the evaluation of Eq.~(\ref{perturb_A}) we apply Wick's theorem
with appropriate propagators for quarks, mesons and gluons. 

For the quark field we use a Feynman
propagator for a fermion in a binding potential with
\eq\label{quark_propagator}
 i G_\psi(x,y) = \langle 0|T\{\psi(x) \bar\psi(y)\}|0 \rangle
\en 
and 
\eq
i G_\psi(x,y) 
&=& \theta(x_0-y_0) \, \sum\limits_{\alpha} \, 
     u_\alpha(\vec{x}) \, \bar u_\alpha(\vec{y}) \, e^{-i{\cal E}_\alpha (x_0-y_0)}\nonumber\\
&-& \theta(y_0-x_0) \, \sum\limits_{\beta} \, 
     v_\beta(\vec{x})  \, \bar v_\beta(\vec{y})  \, e^{ i{\cal E}_\beta (x_0-y_0)} \,.
\en
In the present study we restrict the expansion of the quark propagator 
to the ground state both for quarks and antiquarks. 
For the meson fields we adopt the free Feynman propagator with
\eq
i\Delta_{PP^\prime}(x-y) = 
\langle 0|T\{\Phi_P(x)\Phi_{P^\prime}(y)\}|0 \rangle = 
\delta_{PP^\prime}\int\frac{d^4k}{(2\pi)^4i}
\frac{\exp[-ik(x-y)]}{M_P^2 - k^2 - i\epsilon}. 
\en
For the gluon field we use the dressed propagator containing an 
effective quark-gluon coupling constant 
$\alpha_s(k^2) = g_s^2(k^2)/(4\pi)$ 
with a nontrivial momentum dependence. In our considerations we work in
Coulomb gauge to separate the contributions of Coulomb 
($A_0^a$) and transverse ($A_i^a$) gluons in the propagator: 
\begin{eqnarray}\label{D00}
\hspace*{-.75cm}
i \, \alpha_s \, D_{00}^{ab}(x-y) \, = \,   
- \, \delta^{ab} \int\frac{d^4k}{(2\pi)^4i} 
\frac{e^{-ik(x-y)}}{\vec{k}^{\, 2}} \, 
\alpha_s(k^2) 
\end{eqnarray}
and 
\begin{eqnarray}\label{Dij}
\hspace*{-.75cm}
i \, \alpha_s \, D_{ij}^{ab}(x-y) \, = \,  
\delta^{ab} \int\frac{d^4k}{(2\pi)^4i} 
\frac{e^{-ik(x-y)}}{k^2 + i \varepsilon} 
\biggl\{ \delta_{ij} - \frac{k_ik_j}{\vec{k}^{\, 2}} \biggr\} \, 
\alpha_s(k^2) \,.   
\end{eqnarray} 
Following an approach to low-energy QCD as based on 
the solutions of the Dyson-Schwinger equations~\cite{Maris:1999nt} 
we suppose that the running coupling constant $\alpha_s(k^2)$ 
includes nontrivial effects of vertex and self-energy corrections, etc. 
In the present paper we use a simple analytic form for $\alpha_s$ 
as suggested in Ref.~\cite{Maris:1999nt}:  
\begin{eqnarray}
\alpha_s(t) \, = \, \frac{\pi}{\omega^6} D {t}^2 
 e^{- t/\omega^2} + \frac{2 \gamma_m \pi}
{\ln\biggl[\tau + \left(1 + t/\Lambda_{QCD}^2\right)^2 
\biggr]} F(t)
\end{eqnarray}
where $t = -k^2$ is an Euclidean momentum squared,  
$F(t)= 1-\exp(-t/[4m_t^2])$\,, $\tau = e^2-1$\,, 
$\gamma_m = 12/(33 - 2 N_f)$\,,  $\Lambda_{QCD}^{N_f=4}=0.234$ GeV\,, 
$\omega = 0.3$ GeV\, and $m_t = 0.5$ GeV.  
The functional form of $\alpha_s(t)$ was fitted to the perturbative 
QCD result in the ultraviolet region (at large momentum squared)
and is governed by a single parameter $D$ in the infrared region. 
In Ref.~\cite{Maris:1999nt} the parameter $D = (0.884\,\mbox{GeV})^2$ 
was adjusted phenomenologically to reproduce properties of pions and 
kaons described as bound states of constituent quarks. In our 
considerations we fit the effective coupling with 
$D = (0.5\,\mbox{GeV})^2$ such that the $\Delta-N$ mass splitting is 
reproduced~\cite{Inoue:2004}. The running coupling $\alpha_s(t)$ 
between quarks and 
gluons should be considered as effective, since it depends on the way 
its used in phenomenology. Since in our model we utilize gluon as 
well as meson degrees of freedom,
it is natural that our fitted value for $D$ is smaller than the one of
Ref.~\cite{Maris:1999nt}. Namely, the relatively small value for $D$ 
leaves space for meson cloud effects in the infrared region.

\section{Pentaquark baryon systems and mass spectrum}

The pentaquark baryon systems include a four-quark subsystem
which is totally antisymmetric under the permutation of quarks.
We begin with the discussion of this subsystem.
In the context of a potential model, where the residual interaction is treated
perturbatively, it is natural to start with the energetically favored
configuration where all four quarks are in the ground state orbit.
Some models~\cite{Stancu:2003if,Carlson:2003wc}
discuss a configuration with an excited quark which is
the energetically lowest one.
However, the issue of level reversing in the basic five-quark
configuration strongly depends on the dynamics involved in the model. 
In the context of our model we assume no excitation for the beginning.
We will return to this point later on.
With this condition the four-quark subsystem is symmetric in orbital space
and in a color $3_c$ state such that the total pentaquark system
is a color singlet; hence the subsystem is mixed symmetric in color space.
The $\Theta$ pentaquark includes four u and d quarks.
The spin and isospin coupling of the four non-strange quarks can in general
result in
total spin S=0,1,2 and isospin I=0,1,2. The (S=2, I=2) state
is forbidden here, since both the S=2 and I=2 configurations are 
symmetric in the respective space and a totally antisymmetric
four-quark subsystem cannot be formed, when combined
with the orbital and color symmetries.
The possible spin-isospin combinations which result in a totally 
antisymmetric configuration are (S=0, I=1), (S=1, I=0), (S=1, I=2) and 
(S=2, I=1)~\cite{Wybourne:2003if}.

Capstick \etal studied the (S=1, I=2) combination~\cite{Capstick:2003iq}. 
In this model the $\Theta^+$ baryon is a member of
a isotensor multiplet: \{$\Theta^{+++}$, $\Theta^{++}$, $\Theta^{+}$, 
$\Theta^{0}$, $\Theta^{-}$\} and its narrow decay width is explained 
by isospin conservation.
Carlson \etal originally studied the case of (S=1, I=0)~\cite{Carlson:2003pn}.
Yet the authors revised their work by referring to a
configuration with one orbital excitation, in line with their
estimate of the energy-shift due to the specific residual
interaction~\cite{Carlson:2003wc}.
There is experimental indication that the $\Theta^+$ baryon seems to be
an isoscalar~\cite{Stepanyan:2003qr}.
We follow this suggestion and choose the (S=1, I=0) combination in our considerations. 
All four-quark combinations exist simultaneously in the 
present quark model, and each of these will correspond to a
certain baryon multiplet. A complete case for pentaquarks should be able
to put forward a consistent dynamical mechanism to explain the observable part
of these multiplets and the decay characteristics of the experimentally
accessible states.
We will comment on this point later on.

The isosinglet four-quark combination with the $\bf[22]$ Young tableau
in the two-flavor picture
becomes the $\bar 6_f$ in the flavor SU(3) generalization.
Therefore, by adding a single antiquark ($\bar 3_f$) to the four-quark 
system we generate pentaquark states which split into the $8_f$ and 
the $\bar{10}_f$ flavor SU(3) multiplets as in the 
Jaffe-Wilczek~\cite{Jaffe:2003sg} and 
Karliner-Lipkin~\cite{Karliner:2003sy} models. 
The $\Theta^+$ baryon then belongs to the antidecuplet $\bar{10}_f$.
  
The spin-parity of the four-quark subsystem with (S=1, I=0) is $1^+$.
In the context of the model
it is natural to assume that the antiquark is also in a S-wave ground 
state. As a consequence the spin-parity of the 
pentaquark states are $1/2^-$ and $3/2^-$, where, since
no angular-momentum excitations are introduced, the parity is
negative. 
This point is characteristic of the present approach when
compared to other models (see discussion in Section~\ref{Sect_Intro}).
In total we have 36 pentaquark states
distributed among four multiplets: two octets and two antidecuplets 
with two possible spin-parities, $1/2^-$ and $3/2^-$. 
Their explicit wave functions can, for example, be constructed
by dividing the four-quark subsystem into two-quark pairs 
(for a detailed discussion see Ref.~\cite{Carlson:2003pn}). 

In the following we restrict the application of the PCQM to
the study of the mass spectrum of the negative 
parity antidecuplet $\bar{10}_f$.
In the isospin symmetry limit we consider the masses of 
the $\Theta$, N, $\Sigma$ and $\Xi$ baryons denoting the isospin 
multiplets in each row of the antidecuplet.
At this point we do not introduce 
a mixing between the $8_f$ and the $\bar{10}_f$. The resulting masses 
of N and $\Sigma$ may therefore not correspond to physical ones even if the
present approach is realistic.
If and how strong the ensuing mixing develops dynamically is currently
an open issue.
The masses of the $\Theta$ and the $\Xi$ are not affected by this
mixing.
We consider both the spin $1/2$ and $3/2$ antidecuplets and
study the mass splittings, in particular of $\Theta$ and $\Xi$,
induced by the flavor dependent current quark masses and by residual interactions.

First, on the level of the model confinement
the exotic baryon masses of the four different isomultiplets 
develop mass differences due to the different strange quark/antiquark 
content including the hidden one.
In the present approach finite current quark masses contribute 
a linear term with $\sum\limits_{i=1}^5 \, m_i \gamma$ to the baryon mass,
where $\gamma = \int\!d^3 x \, \bar u_0(x) u_0(x)$ is the relativistic reduction factor\cite{Lyubovitskij:2001nm}.
For the number of valence quarks, that is nonstrange ($n$) and strange
quarks ($s$) and similarly for the antiquarks ($\bar n$ and $\bar s$)
in each baryon,
one can use the numbers listed in Table~\ref{tbl:A11} of the Appendix.
In the current model the total number of $s$ plus $\bar s$ is one unit
larger in $\Xi$ than in $\Theta$ as is known from Ref.~\cite{Carlson:2003pn,Close:2003tv}.

\begin{figure}[tb]
\begin{center}
\includegraphics[width=8cm]{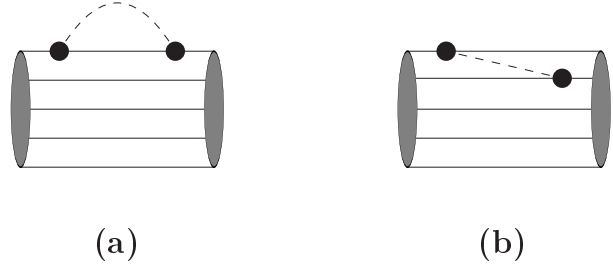}
\end{center}
\caption{\label{fig:m_loop}Meson loop diagrams contributing to the pentaquark
         energy shift: meson cloud (2a) and exchange diagrams (2b).}
\end{figure}

\begin{figure}[tb]
\begin{center}
\includegraphics[width=8cm]{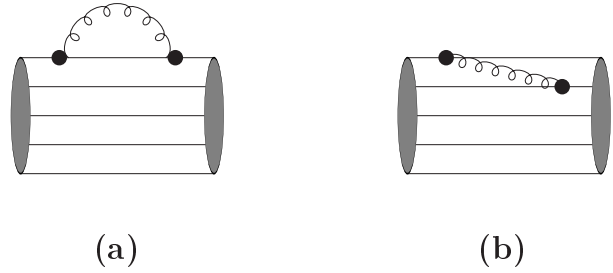}
\end{center}
\caption{\label{fig:g_loop}Gluon loop diagrams contributing to the pentaquark
         energy shift: gluon cloud (3a) and exchange diagrams (3b).}
\end{figure}

Second, the residual quark interaction induces energy shifts
depending on the particular pentaquark state.
In the PCQM the energy shift of the pentaquark valence particles interacting with 
pseudoscalar mesons and quantum gluon fields
is evaluated perturbatively as
\begin{equation}
 \Delta m_B  =
 {}^B\!\langle \phi_0| \sum_{n=1}^{2} \frac{i^n}{n!} 
  \! \int \! i \delta(t_1) d^4 x_1 \ldots d^4 x_n  
 T[{\cal L}_I (x_1) \ldots {\cal L}_I(x_n) ]
 | \phi_0 \rangle^B_c 
\end{equation}
where ${\cal L}_I(x)$ is the interaction Lagrangian set up in
Eq. (\ref{linearized_L}).
The resulting diagrams contained in $\Delta m_B$ are 
shown in Fig.~\ref{fig:m_loop} (meson contribution)
and in Fig.~\ref{fig:g_loop} (gluon contribution) where the five 
internal solid lines represent both the valence quarks and antiquark.
For example, the contribution of pion-exchange between two 
quarks to the pentaquark mass shift is given by 
\begin{equation}\label{eqn:dmpiexqq}
 \Delta m_B^{\pi{\rm -EX};qq} =
  - \frac{1}{6 \pi^2 F^2} \int\limits_0^\infty \!\! d k \, 
  \frac{k^2 \, G^2_{\pi qq}(k^2)}{M_{\pi}^2 + k^2}
  \,  
  \mate<B|\,\displaystyle{\sum_{i<j}^4} \displaystyle{\sum_{a=1}^3} \, 
         {\lambda^{(a)}_i} {\lambda^{(a)}_j} \, 
         \vec \sigma_i \cdot \vec \sigma_j \,|B> \, ,
\end{equation}
where $G_{\pi qq}(k^2)$ is the $\pi qq$ coupling form factor given by:
\begin{equation}
  G_{\pi qq}(k^2) = \int\limits_0^\infty \!\! dr \, r^2 
\left[ 2 f(r) g(r) \right] S(r) j_1(kr) ~.
\end{equation}
Similarly, the contribution of pion-exchange, now between  
a quark and an antiquark, is
\begin{equation}
 \Delta m_B^{\pi{\rm -EX};q\bar q} =
  - \frac{1}{6 \pi^2 F^2} \int\limits_0^\infty \!\! d k \, 
\frac{k^2 \, G_{\pi qq}(k^2)
  G_{\pi \bar q \bar q}(k^2)}{M_{\pi}^2 + k^2}
  \,  
  \mate<B|\,\displaystyle{\sum_{i=1}^4} \displaystyle{\sum_{a=1}^3} \, 
         {\lambda^{(a)}_i} {\lambda^{(a)}_5} \, 
         \vec \sigma_i \cdot \vec \sigma_5 \,|B>
\end{equation}
where index "5'' refers to the antiquark. 
The form factor $G_{\pi \bar q \bar q}(k^2)$
describing the coupling of a $\pi$-meson to two antiquarks
is expressed by the components of the antiquark wave function $v_0(\vec r)$: 
\begin{equation}
  G_{\pi \bar q \bar q}(k^2) = \int\limits_0^\infty \!\! dr \, r^2 
\left[ 2 k(r) l(r) \right] S(r) j_1(kr) ~.
\end{equation}
The contribution of gluon exchange between two quarks 
to the pentaquark mass shift is 
\begin{eqnarray}
 \Delta m_B^{{\rm gl-EX};qq}
  &=&
  \frac{1}{2\pi} \int \limits_0^\infty\!\! dk \, 
 \alpha_s(k^2)\, 
  G_E^2(k^2) \, 
  \mate<B|\,\displaystyle{\sum_{i<j}^4} \, 
         {\vec{\lambda}^C_i} \cdot {\vec{\lambda}^C_j} \,|B>
  \nonumber \\ 
  &-& 
  \frac{1}{3\pi} \int\limits_0^\infty \!\! dk \, 
  \alpha_s(k^2)\, G_M^2(k^2) \, 
  \mate< B|\, \displaystyle{\sum_{i<j}^4} \, 
          {\vec{\lambda}^C_i} \cdot {\vec{\lambda}^C_j} \, 
          \vec \sigma_i \cdot \vec \sigma_j \,|B>
\label{eqn:dmgluon}
\end{eqnarray}
where the first and second terms arise from the electric and magnetic contribution,
respectively. The corresponding form factors $G_E(k^2)$ and $G_M(k^2)$ are 
defined by
\begin{eqnarray}
  G_E(k^2) &=& \int\limits_0^\infty\!\! dr \, r^2 
  \left[ f^2(r) + g^2(r)\right] j_0(kr) \\
  G_M(k^2) &=& \int\limits_0^\infty\!\! dr \, r^2 
  \left[ 2 f(r) g(r)    \right] j_1(kr),
\end{eqnarray}
where $\alpha_s(k^2)$ is the running quark-gluon coupling constant 
(see Section~\ref{Sect_PCQM}). 

The contribution of all other occurring diagrams can be given in a similar
fashion.
We omit so called Z-type self-energy Feynman diagrams to 
avoid possible double counting. 
All relevant flavor-spin and color-spin 
matrix elements are summarized in the Appendix. 

\begin{table}[tb]
\caption{\label{tbl:1}Mass shifts of the spin 1/2 antidecuplet baryons in units of MeV.}
\begin{center}
\begin{tabular}{ccrcrcr}
 \hline
          &  &  quark mass     & &  meson        & &  gluon  \\
 \hline
 $\Theta$ &  &    $148$        & &  $-429$       & &  $-603$ \\      
 \hline
  N       &  &    $189$        & &  $-399$       & &  $-603$ \\     
 \hline 
 $\Sigma$ &  &    $230$        & &  $-367$       & &  $-603$ \\      
 \hline 
 $\Xi$    &  &    $271$        & &  $-333$       & &  $-603$ \\      
 \hline
\end{tabular}
\end{center}
\end{table}

\begin{table}[tb]
\caption{\label{tbl:2}Mass shifts of the spin 3/2 antidecuplet baryons in units of MeV.}
\begin{center}
\begin{tabular}{ccrcrcr}
 \hline
          &  &     quark mass  & &  meson        & &  gluon  \\
 \hline
 $\Theta$ &  &    $148$        & &  $-450$       & &  $-397$ \\      
 \hline
  N       &  &    $189$        & &  $-423$       & &  $-397$ \\     
 \hline 
 $\Sigma$ &  &    $230$        & &  $-397$       & &  $-397$ \\      
 \hline 
 $\Xi$    &  &    $271$        & &  $-373$       & &  $-397$ \\      
 \hline 
\end{tabular}
\end{center}
\end{table}

Numerical values for the mass shifts of the pentaquark states
are deduced for our specific confining trial potential.
In Tables \ref{tbl:1} and \ref{tbl:2} 
we summarize the results for the mass shifts of the spin 1/2 and 3/2
antidecuplet baryons, respectively. Column "quark mass" refers to the
mass shifts induced by the current quark mass. Including relativistic effects the
current quark mass induces 41 MeV splittings between consecutive rows of the 
antidecuplet. Therefore, because of the quark mass term the $\Xi$ becomes 123 MeV
heavier than the $\Theta$, which is a typical result of 
pentaquark models.

The meson cloud leads to a considerable lowering of the 
antidecuplet baryon masses of about 400 MeV. This value is larger than 
for the case of conventional baryons of about 300 MeV.
SU(3) flavor symmetry breaking as induced by the flavor
dependent meson masses has qualitatively
the same effect as for the case of the current quark mass.
The respective splittings lead to an increase in mass when
descending in the rows of the multiplet.
For example, the $\Xi$ becomes 
heavier than the $\Theta^+$ by 96 MeV in the spin 1/2 antidecuplet,
and by 86 MeV in the spin 3/2 antidecuplet.
The negative mass shift due to the meson cloud is stronger in the
spin 3/2 antidecuplet than in the spin 1/2 one. The difference arises from
the exchange contribution between the quarks and the antiquark. 
This particular contribution is included in some studies \cite{Carlson:2003pn},
while omitted in others \cite{Jennings:2003wz}.
It is repulsive in the total spin 1/2 channel and attractive for the spin 3/2
pentaquark baryons (see Tables \ref{tbl:A2} and \ref{tbl:A3} in the Appendix).
Therefore, a residual quark interaction solely based on a mesonic mechanism
leads to spin 3/2 pentaquark baryons
which are lighter than the flavor partners of the spin 1/2 one.
But in our model the size of the splitting is not so large,
about 20 $\sim$ 40 MeV.
For the usual ground state baryons consisting of three valence
quarks, meson cloud effects lead in the present model to a spin 3/2 decuplet
which is about 100 MeV heavier than the spin 1/2 octet \cite{Inoue:2004}. 

The applied semi-perturbative gluon mechanism 
splits the two spin states by 206 MeV, such that the spin 3/2 antidecuplet 
is heavier than the spin 1/2 one. 
This feature is quite similar to that of the ground state spin 1/2 and 3/2 case,
where three valence quarks are involved. 
The origin of the splitting is traced to the magnetic part of the
exchange contribution between the quarks 
and the antiquark. For the spin 1/2 pentaquark baryon it is attractive 
while for spin 3/2 it is repulsive (see Tables~\ref{tbl:A4} 
and \ref{tbl:A5} in the Appendix).
Here, we use the same dressed gluon propagator for the exchange between two quarks, 
and also between a quark and an antiquark.
The results we obtain for the gluon induced mass shifts
are qualitatively consistent with Ref.\cite{Jennings:2003wz}.

Within the hybrid mechanism of the present model,
the spin 3/2 antidecuplet baryons become in total heavier than the spin 1/2 one.
Within this context the observed $\Theta^+$ is likely to be assigned to
the lighter spin 1/2 multiplet.
The splitting of about 185 MeV, however, is not so large.
As a consequence of the present scenario the
spin 3/2 $\Theta$ baryon is located at 1725 MeV.
The negative parity spin 1/2 $\Theta$ at around 1540 MeV can decay
to the S-wave KN system, while for spin 3/2 the $\Theta$ baryon cannot.
In this scenario we still need to account for the observed narrow decay width
of the $\Theta$ baryon.
The reduced overlap between the decay channel KN and the $\Theta$ resonance
with the present color-spin-flavor wave function
provides a suppression of the decay width by a factor 1/4 \cite{Carlson:2003pn}.
This can explain part of the weak coupling to the KN channel.

The mass difference between the $\Theta$ and the $\Xi$ in the spin 1/2 
antidecuplet is 220 MeV in total. Using a $\Theta$ mass of 1540 MeV
as input the model predicts a spin 1/2 $\Xi$ baryon with negative parity at around 1760 MeV.
This result is 100 MeV lower than the recent experimental finding for 
a $\Xi^{--}$ at about 1860 MeV~\cite{Alt:2003vb}.
In the present study we have used a simplified confinement model
leading to symmetric quarks/antiquark wave functions.
The discrepancy may indicate that a more realistic confinement is needed.
At least in the context of the mass spectrum the two possibly
observed $\Theta^+$ and $\Xi^{--}$ baryons can be associated
with members of the $1/2¯$ multiplet.
In the present model effects of flavor symmetry breaking work
constructively and lead to considerable splitting between these two states.
A discussion of only the mass spectrum is obviously not sufficient
to judge the nature of the observed exotic baryon states.
In particular, experimental information on parity and spin of
both states is urgently needed to constrain and also select
the theoretical approaches.
 
\begin{table}[htbp]
\caption{\label{tbl:3} Antidecuplet pentaquark masses in units of MeV.}
\begin{center}
\begin{tabular}{ccccc}
 \hline  
               & $\Theta$ &    N      &  $\Sigma$ &  $\Xi$   \\
 \hline  
 $J^P= 1/2^-$  & $1540$   &   $1611$  &   $1684$  &  $1759$  \\
 \hline  
 $J^P= 3/2^-$  & $1725$   &   $1793$  &   $1860$  &  $1925$  \\
 \hline  
\end{tabular}
\end{center}
\end{table}

In the present simple model the energy of an unperturbed single quark or
antiquark in the ground state orbit is $537 \,\mbox{MeV} \pm V_0$, respectively, 
where $V_0$ is the value of the constant vector potential 
(see Section~\ref{Sect_PCQM}). Starting from this independent
particle model the unperturbed mass of the 
pentaquark system is given by $2685 \,\mbox{MeV} + 3 V_0 - E_{\rm cm}^{\rm 5q}$,
where $E_{\rm cm}^{\rm 5q}$ represents the spurious energy of the center-of-mass
motion.
By simply setting $3 V_0 - E_{\rm cm}^{\rm 5q} = -261$ MeV and using
in addition the obtained total 
mass shift of $-884$ MeV for the spin 1/2 $\Theta$, 
we fit the mass of the $\Theta^+$ baryon of 1540 MeV.  
Similarly, the unperturbed mass of the three-quark system is given by 
$1611 \,\mbox{MeV} + 3 V_0 - E_{\rm cm}^{\rm 3q}$, and in this case
the nucleon mass can be fitted as well.
The two spurious energies $E_{\rm cm}^{\rm 3q}$ and $E_{\rm cm}^{\rm 5q}$
are in general different (judging from the non-relativistic
harmonic oscillator potential $E_{\rm cm}^{\rm 5q}$ should be
smaller than $E_{\rm cm}^{\rm 3q}$), hence
the mass difference between the nucleon and the $\Theta^+$ cannot be
predicted in the present approach.
The obtained mass spectrum of antidecuplet pentaquarks is indicated in
Table\ref{tbl:3}.

Finally, we give indications for the validity of our
basic model configuration, where all valence quarks and the antiquark
are placed in ground state orbits with the four-quark
coupling (S=1, T=0).
In the work by 
Carlson \etal~\cite{Carlson:2003wc} it is shown that a configuration 
with one excited quark is energetically favored to the one without, 
despite the additional excitation energy under the influence of the
spin-isospin dependent interquark force. This is not the case in our model
although we also have 
a spin-isospin force induced by one-pion exchange. 
In Ref.~\cite{Carlson:2003wc} the strength of this force
is chosen such to reproduce the full N-$\Delta$ splitting, 
and is about 3 times stronger than in our case. 
When repeating their calculation with a reduced force by a factor of 1/3, 
the levels do not reverse anymore. Hence, the level reversing found in 
Ref.~\cite{Carlson:2003wc} does not pose a problem in the
current model context.

At present it is certainly worthwhile to consider other basic
pentaquark configurations (such as the I=1 four-quark combination
and also the ones including single particle excitations)
to obtain the full mass spectra of these systems.
Mixing between these configurations and coupling to the decay
channels furthermore complicate this issue.
Again, a clear dynamical argument should be finally developed
to work out the observable part of the pentaquark spectra in
close relation to present and future experimental findings.

\section{$\Theta^+$ baryon decay width} 

In this Section we discuss a possible mechanism for a small decay width of 
the $\Theta^+$ baryon in the framework of PCQM using an uncorrelated 
five-quark configuration for pentaquarks. Different models have been 
applied to explain an extremely small width of $\Gamma_{\Theta^+}$ 
for the $\Theta^+$ baryon (see detailed discussion in recent 
papers~\cite{Cahn:2003wq,Eidemuller:2005jm,Diakonov:2005ib}). 
The current experimental status of the $\Theta^+$ width as follows. 
Several recent experiments give upper limits in range 
1-4 MeV~\cite{Eidemuller:2005jm}. 
The experiment on $K^+$ collisions on xenon and deuterium gives 
smaller value of $\Gamma_{\Theta^+} = 0.9 \pm 0.3$ MeV~\cite{Cahn:2003wq}.  
Below we derive the expression for width of $\Theta^+(\frac{1}{2}^-)$ baryon 
in terms of unknown coupling of the kaon to the strange antiquark and three 
nonstrange quarks. Next we estimate this coupling using our formalism. 

\begin{figure}[tb]
\begin{center}
\includegraphics[width=10cm]{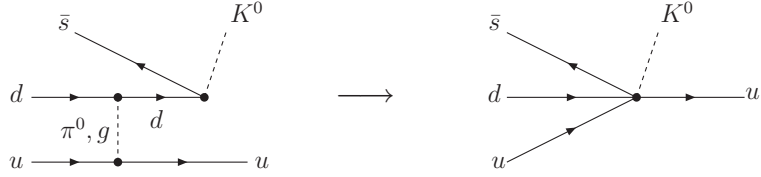}
\caption{\label{fig:pentaq}Possible mechanism to generate 
effective coupling $K^0 \bar s u u d$.} 
\end{center}
\end{figure}

\begin{figure}[tb]
\begin{center}
\includegraphics[width=8cm]{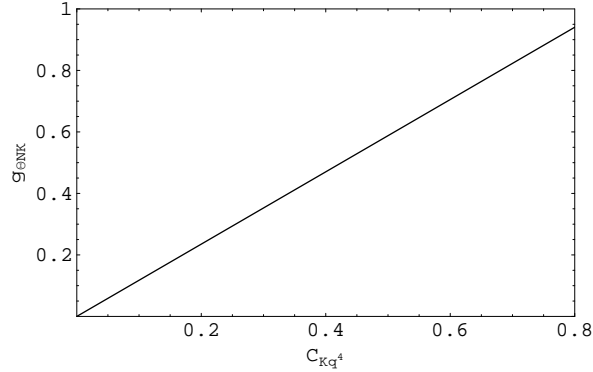}
\end{center}
\caption{\label{fig:coupling}Coupling $g_{\Theta NK}$ 
as function of $c_{Kq^4}$.} 
\end{figure}

\begin{figure}[tb]
\begin{center}
\includegraphics[width=8cm]{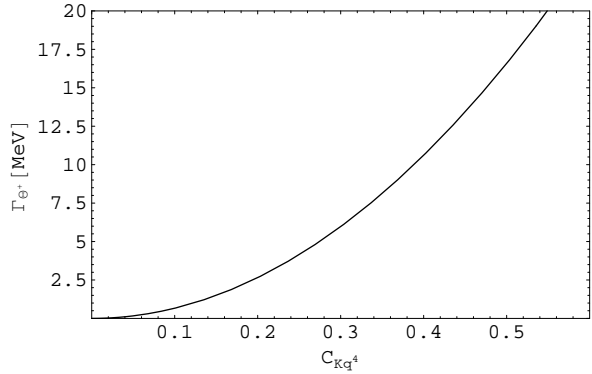}
\end{center}
\caption{\label{fig:gamma}Width $\Gamma_{\Theta^+}$ 
as function of $c_{Kq^4}$.} 
\end{figure}

To generate a connected Feynman diagram describing the transition 
$\Theta^+ \to  p (n) + K^0 (K^+)$ we need a coupling of the kaon 
to the strange antiquark and three nonstrange quarks: $K^0 \bar s u u d$ 
for the decay into the proton and $K^+ \bar s u d d$ for the decay 
into the neutron. 
In the framework of the constituent quark model this vertex can be 
understood as an effective coupling generated in second order of 
perturbation theory (for example, due to the exchange by a neutral pion 
or gluon). 
Therefore, we suggest that the transition $\Theta^+ \to  p (n) + K^0 (K^+)$ 
is governed by a higher-order operator which probably explains the small 
width of the $\Theta^+$ baryon. We understand that this scenario is not 
unique and a more detailed study of the problem should be done, but 
it goes beyond the current manuscript. 
In Fig.~\ref{fig:pentaq} we draw a possible second-order diagram (left 
panel) generating the effective $K^0 \bar s u u d$ coupling which 
is a part of the effective local Lagrangian 
\eq\label{LKq4} 
{\cal L}_{Kq^4}(x) = - \frac{c_{Kq^4}}{F^3} 
         \bar K^0(x) \bar u(x) \Gamma_1 u(x) \bar s(x) \Gamma_2 d(x) 
+ {\rm h.c.}  
\en 
where $c_{Kq^4}$ is an unknown dimensionless coupling constant; 
$\Gamma_1$ and $\Gamma_2$ are the Dirac spin matrices. The coupling 
$K^+ \bar s u d d$ (diagram and effective Lagrangian) can be obtained via 
the replacements $u \leftrightarrow d$ and $K^0 \to K^+$. 
We choose the simplest set of possible 
spin matrices in the Lagrangian (\ref{LKq4}): $\Gamma_1 = I$ and 
$\Gamma_2 = i \gamma_5$. The effective coupling 
$g_{\Theta NK}$ for the negative-parity pentaquark is defined as 
\eq 
{\cal L}_{\Theta NK}(x) \, = \, i \, g_{\Theta NK} \, 
\bar N(x) \, \Theta^+(x) \, K(x) \, + \, {\rm h.c.}  
\en 
where $N$ and $K$ are the doublets of nucleons $(p,n)$ and kaons $(K^0,K^+)$, 
respectively. A straightforward calculation of $g_{\Theta NK}$ in the PCQM 
relates it to $c_{Kq^4}$: 
\eq 
g_{\Theta NK} \, = \, \frac{1}{32\pi^2} \, \frac{c_{Kq^4}}{F^3}  \, 
\int d^3 x \, [g^2(x) - f^2(x)]  \, [g^2(x) + f^2(x)] 
\en   
where $g$ and $f$ are the components of the ground-state quark wave 
function~(\ref{uv_0}). On the other hand, $g_{\Theta NK}$ is related 
to the $\Theta^+$ width as~\cite{Eidemuller:2005jm} 
\eq 
\Gamma_{\Theta^+} \, = \, \frac{g_{\Theta NK}^2}{8\pi m_\Theta^3} \, 
\lambda^{1/2}(m_\Theta^2,m_N^2,M_K^2) \, [(m_\Theta + m_N)^2 - M_K^2]  
\en  
where $\lambda(x,y,z) = x^2 + y^2 + z^2 - 2xy - 2yz - 2xz$ 
is the K\"allen triangle function. 
In Figs.~\ref{fig:coupling} and~\ref{fig:gamma} we demonstrate the 
behavior of the quantities $g_{\Theta NK}$ and $\Gamma_{\Theta^+}$ 
as functions of effective constant $c_{Kq^4}$. Note, that 
the central value of $\Gamma_{\Theta^+} = 0.9 \pm 0.3$~MeV~\cite{Cahn:2003wq}
corresponds to $c_{Kq^4} = 0.12$. Using our formalism we can estimate  
the coupling $c_{Kq^4}$ relying on the idea that the two-body forces between 
nonstrange quarks are generated by the one-pion exchange (see 
left panel in Fig.4). Our result is: 
$c_{Kq^4} = 0.33$. It gives $\Gamma_{\Theta^+} = 7$ MeV which overestimates 
the current upper limits for this quantity~\cite{Cahn:2003wq}.  
Other possibilities for the 
negative-parity $\Theta^+$ pentaquark to have a small width is also 
discussed in the content of other theoretical approaches (see, 
for example, discussion in Refs.~\cite{Eidemuller:2005jm,Wang:2005ms}). 

\section{Summary}

In this paper, we have studied mass spectrum of the $J^P=1/2^-$ and 
$3/2^-$ pentaquark antidecuplets initiated by the newly discovered 
exotic $\Theta^+$ baryon. We have applied the perturbative chiral quark 
model, where confined valence quarks/antiquarks interact with meson 
fields according to chiral symmetry requirements and with quantum 
fluctuations of the gluon field. Guided by experimental preference
we selected a specific configuration of 
the pentaquark system leading to negative parity baryon multiplets:
the flavor SU(3) octet and antidecuplet with spin 1/2 and 3/2 for each
of these multiplets. 
The $\Theta^+$ baryon can be considered as a possible member of one
of the antidecuplets.
Here we payed attention to the antidecuplet only,
and evaluated the energy shifts arising from the residual interaction as well as 
from the current quark masses.
The model parameters, \ie the confining potential 
and the effective quark-gluon coupling,
are set up and constrained such as to give a
reasonable fit to mass shifts in the octet and decuplet sector of conventional
baryons.

Within the model we generate a spin 1/2 antidecuplet which
is lighter than for spin 3/2.
The observed $\Theta^+$ baryon can be assigned as a member of the spin 1/2 
antidecuplet with negative parity. The origin of the splitting between
the two multiplets is dominantly traced to the 
semi-perturbative gluon exchange between the quarks and the antiquark.
The size of the splitting is about 185 MeV, which is somewhat smaller
when compared to the usual mass difference of the
conventional spin 1/2 and 3/2 ground state baryons.
This qualitative difference arises from the meson exchange contribution,
which shifts the
spin 3/2 pentaquark antidecuplet slightly lower with respect to the
spin 1/2 one,
while the spin 3/2 decuplet of the ground state baryons becomes
about 100 MeV heavier relative to the octet.
The two spin states are a necessary consequence in most of
the pentaquark models, however the splitting between these two are often not studied. 

The current quark mass splits the baryon mass inside each 
antidecuplet according to its strange quark/antiquark content.
With the present relativistic quark wave function,
the size of the splitting is about 41 MeV between each neighboring row
of the isospin multiplets.
By this mechanism the bottom member $\Xi$ becomes 123 MeV heavier than the
$\Theta$. In addition, the inclusion of the meson cloud gives a considerable 
mass splitting within each antidecuplet.
For instance, we obtain a mass difference of 96 MeV between the spin 1/2
states $\Theta$ and $\Xi$ due to meson loops.
Our results for SU(3) flavor violation within the multiplets are
originally induced by flavor dependent quark masses, 
which in turn affect the meson masses, and thereby meson loops.

As a final result we have a spin 1/2 $\Xi$ which is 220 MeV heavier
than its $\Theta$ partner in the multiplet.
This value is 30\% smaller when compared to 
the preliminary mass difference of the $\Theta^+$ and $\Xi^{--}$
deduced from data. This result may indicate that the
present model is reasonable but probably too simplified in
treating confinement. We have not studied the decay widths of 
the pentaquark system yet. The observed narrow decay width of the 
$\Theta^+$ baryon is a key challenge to explain. At the same
time the possible observability of the other members
of the multiplet should be worked out in consistency
with experimental findings.
The issue of the decay patterns of the antidecuplet pentaquark states
will be subject of a forthcoming paper.

Finally, we discuss a possible mechanism for a small decay 
width of the negative-parity $\Theta^+$ baryon. We suggest 
an existence of relevant effective transition operator which 
can be generated in the second-order of perturbation theory. 
Our result is: $\Gamma_{\Theta^+} = 7$ MeV 
which overestimates the current upper limits for this 
quantity: $\Gamma_{\Theta^+} = 0.9 \pm 0.3$ MeV~\cite{Cahn:2003wq}.  

\section*{acknowledgments}
This work was supported by the Deutsche Forschungsgemeinschaft (DFG) 
under contracts FA67/25-3 and GRK683. This research is also part of 
the EU Integrated Infrastructure Initiative Hadron physics project 
under contract number RII3-CT-2004-506078 and  
President grant of Russia "Scientific Schools"  No. 1743.2003. 

\appendix
\section{Flavor-spin and color-spin matrix elements}

 In the following we list the flavor-spin and color-spin matrix elements 
 of the antidecuplet pentaquark baryons discussed in the text.
 Here, the particle label 5 refers to the antiquark while 1 to 4 to
 the valence quarks.
 The symbols $\lambda^{(a=1 \sim 8)}$ are 
 the flavor Gell-Mann matrices,
 $\vec \sigma$ is the vector of the Pauli spin matrices, 
 and $\vec \lambda^C$ is the vector of the color Gell-Mann matrices.
 In some matrix elements partial contributions of the non-strange quark($n$),
 the strange quark ($s$), the non-strange ($\bar n$) and strange 
 antiquark ($\bar s$) are also given.

\begin{table}
\caption{\label{tbl:A1}Matrix element 
  $\mate<1^C\!,\bar{10}^F\!,\frac12^{S}(\frac32^{S})| 
         \displaystyle{\sum_{i<j}^4} \displaystyle{\sum_{a}} 
         {\lambda^{(a)}_i} {\lambda^{(a)}_j} 
         \vec \sigma_i \cdot \vec \sigma_j 
        |1^C\!,\bar{10}^F\!,\frac12^{S}(\frac32^{S})>$
   in units of 1/27.} 
\begin{center}
\begin{tabular}{ccrc|rc|rrrr}
\hline
& & $a=1 \sim 3$ & & $a=4 \sim 7$ 
& & \multicolumn{4}{c}{$a=8$}          \\
& &    $nn$      & &   $ns$       & &   $nn$   
&  $ns$  &  $ss$  & total \\
 \hline
$\Theta$& &   $270$      & &    $0$       & &   $-18$  
&   $0$ & $0$  & $-18$ \\
 \hline
N       & &   $180$      & &   $72$       & &   $-12$  
&  $12$  &   $0$  &   $0$ \\
 \hline 
$\Sigma$& &    $93$      & &  $132$       & &    $-5$  
&  $28$  &   $4$  &  $27$ \\
 \hline 
$\Xi$   & &     $9$      & &  $180$       & &     $3$  
&  $48$  &  $12$  &  $63$ \\
\hline
\end{tabular}
\end{center}
\end{table}

\begin{table}
\caption{\label{tbl:A2}Matrix element
  $\mate<1^C\!,\bar{10}^F\!,\frac12^{S}|
         \displaystyle{\sum_{i=1}^4} \displaystyle{\sum_{a}} \, 
         {\lambda^{(a)}_i} {\lambda^{(a)}_5} \, 
         \vec \sigma_i \cdot \vec \sigma_5 \,
        |1^C\!,\bar{10}^F\!,\frac12^{S}>$
  in units of 1/9.} 
\begin{center}
\begin{tabular}{ccrc|rc|rrrrr}
\hline
& & $a=1 \sim 3$ & & $a=4 \sim 7$         
& & \multicolumn{5}{c}{$a=8$}  \\
& &  $n\bar n$   & & $n\bar n$($s\bar s$) & & $n\bar n$ 
& $n\bar s$ & $s\bar n$ & $s\bar s$ & total\\
 \hline
$\Theta$ & &   $0$  & &    $0$  & &   $0$ & $-24$ &   $0$ & $0$ &$-24$ \\
 \hline
N       & &   $0$  & &  $-24$  & &   $4$ & $-12$ &   $0$ & $8$ &  $0$ \\
 \hline 
$\Sigma$ & &  $-6$  & &  $-24$  & &   $6$ &  $-4$ &  $-4$ & $8$ &  $6$ \\
 \hline 
$\Xi$    & & $-18$  & &    $0$  & &   $6$ &   $0$ & $-12$ & $0$ & $-6$ \\
\hline
\end{tabular}
\end{center}
\end{table}

\begin{table}
\caption{\label{tbl:A3}Matrix element
  $\mate<1^C\!,\bar{10}^F\!,\frac32^{S}|
         \displaystyle{\sum_{i=1}^4} \displaystyle{\sum_{a}} \,
         {\lambda^{(a)}_i} {\lambda^{(a)}_5} \, 
         \vec \sigma_i \cdot \vec \sigma_5 \,
        |1^C\!,\bar{10}^F\!,\frac32^{S}>$
  in units of 1/9.} 
\begin{center}
\begin{tabular}{ccrc|rc|rrrrr}
\hline
& & $a=1 \sim 3$ & & $a=4 \sim 7$         
& & \multicolumn{5}{c}{$a=8$}  \\
& &  $n\bar n$   & & $n\bar n$($s\bar s$) & & $n\bar n$ 
& $n\bar s$ & $s\bar n$ & $s\bar s$ & total\\
 \hline
 $\Theta$ & & $0$  & &   $0$ & &  $0$   &  $12$  &  $0$ &  $0$ & $12$ \\
 \hline
  N       & & $0$  & &  $12$ & &  $-2$  &   $6$  &  $0$ & $-4$ &  $0$ \\
 \hline 
 $\Sigma$ & & $3$  & &  $12$ & &  $-3$  &   $2$  &  $2$ & $-4$ & $-3$ \\
 \hline 
 $\Xi$    & & $9$  & &   $0$ & &  $-3$  &   $0$  &  $6$ &  $0$ &  $3$ \\
\hline
\end{tabular}
\end{center}
\end{table}

\begin{table}
\caption{\label{tbl:A4}Matrix element
  $\mate<1^C\!,\bar{10}^F\!,\frac12^{S}(\frac32^{S})|
         \displaystyle{\sum_{i=1}^4} \displaystyle{\sum_{a}} \,
        {\lambda^{(a)}_i} {\lambda^{(a)}_i} \,
        |1^C\!,\bar{10}^F\!,\frac12^{S}(\frac32^{S})>$
  in units of 1/9.} 
\begin{center}
\begin{tabular}{ccrc|rrrc|rrr}
\hline
& & $a=1 \sim 3$  & &  \multicolumn{3}{c}{$a=4 \sim 7$} 
& &  \multicolumn{3}{c}{$a=8$} \\
         & &  $n$  & &  $n$   &  $s$ & total & &   $n$  &  $s$ & total \\
 \hline
$\Theta$ & & $108$ & &  $72$  &  $0$ &  $72$ & &  $12$  &  $0$ & $12$ \\
 \hline
 N       & & $90$  & &  $60$  & $24$ &  $84$ & &  $10$  &  $8$ & $18$ \\
 \hline 
$\Sigma$ & &  $72$ & &  $48$  & $48$ &  $96$ & &   $8$  & $16$ & $24$ \\
 \hline 
$\Xi$    & &  $54$ & &  $36$  & $72$ & $108$ & &   $6$  & $24$ & $30$  \\
\hline
\end{tabular}
\end{center}
\end{table}

\begin{table}
\caption{\label{tbl:A5}Matrix element
  $\mate<1^C\!,\bar{10}^F\!,\frac12^{S}(\frac32^{S})|
  \displaystyle{\sum_{a}} \, {\lambda^{(a)}_5} {\lambda^{(a)}_5} \,
        |1^C\!,\bar{10}^F\!,\frac12^{S}(\frac32^{S})>$
  in units of 1/9.} 
\begin{center}
\begin{tabular}{ccrc|rrrc|rrr}
\hline
& & $a=1 \sim 3$  & &  \multicolumn{3}{c}{$a=4 \sim 7$} 
& &  \multicolumn{3}{c}{$a=8$} \\
& &  $\bar n$  & &  $\bar n$   &  $\bar s$ & total & &   $\bar n$  &  $\bar s$ & total \\
 \hline
$\Theta$ & &  $0$  & &   $0$  & $36$ &  $36$ & &   $0$  & $12$  & $12$ \\
 \hline
N        & &  $9$  & &   $6$  & $24$ &  $30$ & &   $1$  &  $8$  & $9$ \\
 \hline 
$\Sigma$ & & $18$  & &  $12$  & $12$ &  $24$ & &   $2$  &  $4$  & $6$ \\
 \hline 
 $\Xi$   & & $27$  & &  $18$  &  $0$ &  $18$ & &   $3$  &  $0$  & $3$ \\
\hline
\end{tabular}
\end{center}
\end{table}

\begin{table}
\caption{\label{tbl:A6}Matrix element
  $\mate<1^C\!,\bar{10}^F\!,\frac12^{S}(\frac32^{S})| \,
         \displaystyle{\sum_{i<j}^4} \, 
         {\vec{\lambda}^C_i} \cdot {\vec{\lambda}^C_j} \, 
         \vec \sigma_i \cdot \vec \sigma_j \,
        |1^C\!,\bar{10}^F\!,\frac12^{S}(\frac32^{S})>$
  in units of 1/27.} 
\begin{center}
\begin{tabular}{ccrcrcrcr}
\hline
           & &  $nn$  & &  $ns$  & &  $ss$ & & total\\
 \hline
 $\Theta$  & & $144$  & &   $0$  & &   $0$ & & 144 \\
 \hline
  N        & &  $96$  & &  $48$  & &   $0$ & & 144 \\
 \hline 
 $\Sigma$  & &  $22$  & & $148$  & & $-26$ & & 144 \\
 \hline 
 $\Xi$     & & $-78$  & & $300$  & & $-78$ & & 144 \\
\hline
\end{tabular}
\end{center}
\end{table}

\begin{table}
\caption{\label{tbl:A7}Matrix element
  $\mate<1^C\!,\bar{10}^F\!,\frac12^{S}|\,
         \displaystyle{\sum_{i=1}^4} \, 
         {\vec{\lambda}^C_i} \cdot {\vec{\lambda}^C_5} \, 
         \vec \sigma_i \cdot \vec \sigma_5 \,
        |1^C\!,\bar{10}^F\!,\frac12^{S}>$
  in units of 1/9.} 
\begin{center}
\begin{tabular}{ccrcrcrcrcr}
\hline
    & &   $n\bar n$  & &   $n\bar s$ & &   $s \bar n$ & &  $s \bar s$ & & total\\
 \hline
 $\Theta$  & &    $0$  & &   $120$     & &   $0$        & &   $0$  & & 120 \\
 \hline
  N        & &   $40$  & &    $60$     & &   $0$        & &  $20$  & & 120 \\
 \hline 
 $\Sigma$  & &   $60$  & &    $20$     & &  $20$        & &  $20$  & & 120 \\
 \hline 
 $\Xi$     & &   $60$  & &     $0$     & &  $60$        & &   $0$  & & 120 \\
\hline
\end{tabular}
\end{center}
\end{table}

\begin{table}
\caption{\label{tbl:A8}Matrix element
  $\mate<1^C\!,\bar{10}^F\!,\frac32^{S}|\,
         \displaystyle{\sum_{i=1}^4} \, 
         {\vec{\lambda}^C_i} \cdot {\vec{\lambda}^C_5} \, 
         \vec \sigma_i \cdot \vec \sigma_5 \,
        |1^C\!,\bar{10}^F\!,\frac32^{S}>$
  in units of 1/9.} 
\begin{center}
\begin{tabular}{ccrcrcrcrcr}
\hline
            & &  $n\bar n$  & &  $n\bar s$  & &  $s \bar n$ & & $s \bar s$ & & total \\
 \hline
 $\Theta$   & &   $0$  & &    $-60$    & &    $0$      & &   $0$  & & $-60$ \\
 \hline
  N         & & $-20$  & &    $-30$    & &    $0$      & & $-10$  & & $-60$ \\
 \hline 
 $\Sigma$   & & $-30$  & &    $-10$    & &  $-10$      & & $-10$  & & $-60$ \\
 \hline 
 $\Xi$      & & $-30$  & &      $0$    & &  $-30$      & &   $0$  & & $-60$ \\
\hline
\end{tabular}
\end{center}
\end{table}

\begin{table}
\caption{\label{tbl:A9}Matrix element
  $\mate<1^C\!,\bar{10}^F\!,\frac12^{S}(\frac32^{S})|\,
         \displaystyle{\sum_{i<j}^4} \, 
         {\vec{\lambda}^C_i} \cdot {\vec{\lambda}^C_j} \,
        |1^C\!,\bar{10}^F\!,\frac12^{S}(\frac32^{S})>$
  in units of 1/9.} 
\begin{center}
\begin{tabular}{ccrcrcrcr}
\hline
            & &  $nn$  & &  $ns$  & &  $ss$ & & total\\
 \hline
 $\Theta$   & & $-72$  & &  $0$   & &   $0$ & & $-72$ \\
 \hline
  N         & & $-48$  & & $-24$  & &   $0$ & & $-72$ \\
 \hline 
 $\Sigma$   & & $-30$  & & $-36$  & &  $-6$ & & $-72$ \\
 \hline 
 $\Xi$      & & $-18$  & & $-36$  & & $-18$ & & $-72$ \\
\hline
\end{tabular}
\end{center}
\end{table}

\begin{table}
\caption{\label{tbl:A10}Matrix element
  $\mate<1^C\!,\bar{10}^F\!,\frac12^{S}(\frac32^{S})| \,
        \displaystyle{\sum_{i=1}^4} \, 
        {\vec{\lambda}^C_i} \cdot {\vec{\lambda}^C_5} \,
        |1^C\!,\bar{10}^F\!,\frac12^{S}(\frac32^{S})>$
  in units of 1/9.} 
\begin{center}
\begin{tabular}{ccrcrcrcrcr}
\hline
            & &$n\bar n$ & &$n\bar s$ & & $s\bar n$ & & $s\bar s$ & & total \\
 \hline
 $\Theta$   & &    $0$   & &  $-48$   & &    $0$    & &  $0$  & & $-48$ \\
 \hline
  N         & &  $-16$   & &  $-24$   & &    $0$    & & $-8$  & & $-48$ \\
 \hline 
 $\Sigma$   & &  $-24$   & &   $-8$   & &   $-8$    & & $-8$  & & $-48$ \\
 \hline 
 $\Xi$      & &  $-24$   & &    $0$   & &  $-24$    & &  $0$  & & $-48$ \\
\hline
\end{tabular}
\end{center}
\end{table}

\begin{table}
\caption{\label{tbl:A11}Matrix element
  $\mate<1^C\!,\bar{10}^F\!,\frac12^{S}(\frac32^{S})| \,
         \displaystyle{\sum_{i=1}^5} \, 
        {\vec{\lambda}^C_i} \cdot {\vec{\lambda}^C_i} \,
        |1^C\!,\bar{10}^F\!,\frac12^{S}(\frac32^{S})>$
  in units of 16/3.} 
\begin{center}
\begin{tabular}{ccccccccccc}
\hline
           & &    $n$       & &   $s$    & & $\bar n$   & & $\bar s$  & & total \\
 \hline
$\Theta$   & &    $4$       & &   $0$     & &    $0$     & &  $1$      & & 5 \\
 \hline
N          & & $\frac{10}3$ & & $\frac23$ & & $\frac13$  & & $\frac23$ & & 5 \\
 \hline 
$\Sigma$   & & $\frac83$    & & $\frac43$ & & $\frac23$  & & $\frac13$ & & 5 \\
 \hline 
$\Xi$      & &  $2$         & &    $2$    & &  $1$       & & $0$       & & 5 \\
\hline
\end{tabular}
\end{center}
\end{table}


\begin{thebibliography}{99}
%
%\cite{Nakano:2003qx}
\bibitem{Nakano:2003qx}
T.~Nakano {\it et al.}  [LEPS Collaboration],
%``Observation of S = +1 baryon resonance in photo-production 
%from  neutron,''
Phys.\ Rev.\ Lett.\  {\bf 91}, 012002 (2003)
[arXiv:hep-ex/0301020]; 
%%CITATION = HEP-EX 0301020;%%
%
%\cite{Barmin:2003vv}
%\bibitem{Barmin:2003vv}
V.~V.~Barmin {\it et al.}  [DIANA Collaboration],
%``Observation of a baryon resonance with positive strangeness in K+
%collisions with Xe nuclei,''
Phys.\ Atom.\ Nucl.\  {\bf 66}, 1715 (2003)
[Yad.\ Fiz.\  {\bf 66}, 1763 (2003)]
[arXiv:hep-ex/0304040] 
%%CITATION = HEP-EX 0304040;%%
%
%\cite{Stepanyan:2003qr}
\bibitem{Stepanyan:2003qr}
S.~Stepanyan {\it et al.}  [CLAS Collaboration],
%``Observation of an exotic S = +1 baryon in exclusive 
%photoproduction from the deuteron,''
Phys.\ Rev.\ Lett.\  {\bf 91}, 252001 (2003)
[arXiv:hep-ex/0307018]; 
%%CITATION = HEP-EX 0307018;%%
%
%\cite{Barth:2003es}
%\bibitem{Barth:2003es}
J.~Barth {\it et al.}  [SAPHIR Collaboration],
 %``Observation of the positive-strangeness pentaquark 
%Theta+ in photoproduction with the SAPHIR detector at ELSA,''
arXiv:hep-ex/0307083.
%%CITATION = HEP-EX 0307083;%%
%
%\cite{Arndt:2003dm}
\bibitem{Arndt:2003dm}
R.~A.~Arndt, I.~I.~Strakovsky and R.~L.~Workman,
%``K+N scattering data and exotic Z+ resonances,''
arXiv:nucl-th/0311030.
%%CITATION = NUCL-TH 0311030;%%
%
%\cite{Alt:2003vb}
\bibitem{Alt:2003vb}
C.~Alt {\it et al.}  [NA49 Collaboration],
%``Observation of an exotic S = -2, Q = -2 baryon resonance 
%in proton proton collisions at the CERN SPS,''
Phys.\ Rev.\ Lett.\  {\bf 92}, 042003 (2004)
[arXiv:hep-ex/0310014]; 
%%CITATION = HEP-EX 0310014;%%
%
%\cite{Fischer:2004qb}
%\bibitem{Fischer:2004qb}
H.~G.~Fischer and S.~Wenig,
%``Are there S = -2 pentaquarks?,''
Eur.\ Phys.\ J.\ C {\bf 37}, 133 (2004) 
[arXiv:hep-ex/0401014]; 
%%CITATION = HEP-EX 0401014;%%
%
%\cite{Price:2004hr}
%\bibitem{Price:2004hr}
J.~W.~Price, J.~Ducote, J.~Goetz and B.~M.~K.~Nefkens  
[CLAS Collaboration],
%``Photoproduction of the doubly-strange Xi hyperons,''
arXiv:nucl-ex/0402006
%%CITATION = NUCL-EX 0402006;%%
%
%\cite{Diakonov:1997mm}
\bibitem{Diakonov:1997mm}
D.~Diakonov, V.~Petrov and M.~V.~Polyakov,
%``Exotic anti-decuplet of baryons: Prediction from chiral solitons,''
Z.\ Phys.\ A {\bf 359}, 305 (1997)
[arXiv:hep-ph/9703373].
%%CITATION = HEP-PH 9703373;%%
%
%\cite{Cohen:2003yi}
\bibitem{Cohen:2003yi}
T.~D.~Cohen,
 %``Chiral soliton models, large N(c) consistency and the Theta+ exotic
%baryon,''
Phys.\ Lett.\ B {\bf 581}, 175 (2004)
[arXiv:hep-ph/0309111].
%%CITATION = HEP-PH 0309111;%%
%
%\cite{Llanes-Estrada:2003us}
\bibitem{Llanes-Estrada:2003us}
F.~J.~Llanes-Estrada, E.~Oset and V.~Mateu,
%``On the possible nature of the Theta+ as a K pi N bound state,''
Phys.\ Rev.\ C {\bf 69}, 055203 (2004)
[arXiv:nucl-th/0311020].
%%CITATION = NUCL-TH 0311020;%%
%
%\cite{Jaffe:2003sg}
\bibitem{Jaffe:2003sg}
R.~L.~Jaffe and F.~Wilczek,
%``Diquarks and exotic spectroscopy,''
Phys.\ Rev.\ Lett.\  {\bf 91}, 232003 (2003)
[arXiv:hep-ph/0307341].
%%CITATION = HEP-PH 0307341;%%
%
%\cite{Karliner:2003sy}
\bibitem{Karliner:2003sy}
M.~Karliner and H.~J.~Lipkin,
%``The constituent quark model revisited: 
%Quark masses, new predictions for hadron masses and K N pentaquark,''
arXiv:hep-ph/0307243.
%%CITATION = HEP-PH 0307243;%%
%
%\cite{Capstick:2003iq}
\bibitem{Capstick:2003iq}
S.~Capstick, P.~R.~Page and W.~Roberts,
%``Interpretation of the Theta+ as an isotensor resonance 
%with weakly  decaying partners,''
Phys.\ Lett.\ B {\bf 570}, 185 (2003)
[arXiv:hep-ph/0307019].
%%CITATION = HEP-PH 0307019;%%
%
%\cite{Stancu:2003if}
\bibitem{Stancu:2003if}
F.~Stancu and D.~O.~Riska,
%``Stable u u d d anti-s pentaquarks in the constituent quark model,''
Phys.\ Lett.\ B {\bf 575}, 242 (2003)
[arXiv:hep-ph/0307010].
%%CITATION = HEP-PH 0307010;%%
%
%\cite{Carlson:2003pn}
\bibitem{Carlson:2003pn}
C.~E.~Carlson, C.~D.~Carone, H.~J.~Kwee and V.~Nazaryan,
%``Phenomenology of the pentaquark antidecuplet,''
Phys.\ Lett.\ B {\bf 573}, 101 (2003)
[arXiv:hep-ph/0307396].
%%CITATION = HEP-PH 0307396;%%
%
%\cite{Carlson:2003wc}
\bibitem{Carlson:2003wc}
C.~E.~Carlson, C.~D.~Carone, H.~J.~Kwee and V.~Nazaryan,
%``Positive parity pentaquarks pragmatically predicted,''
Phys.\ Lett.\ B {\bf 579}, 52 (2004)
[arXiv:hep-ph/0310038].
%%CITATION = HEP-PH 0310038;%%
%
%\cite{Jennings:2003wz}
\bibitem{Jennings:2003wz}
B.~K.~Jennings and K.~Maltman,
%``Z* resonances: Phenomenology and models,''
Phys.\ Rev.\ D {\bf 69}, 094020 (2004)
[arXiv:hep-ph/0308286].
%%CITATION = HEP-PH 0308286;%%
%
%\cite{Zhu:2003ba}
\bibitem{Zhu:2003ba}
S.~L.~Zhu,
%``Understanding pentaquark states in QCD,''
Phys.\ Rev.\ Lett.\  {\bf 91}, 232002 (2003)
[arXiv:hep-ph/0307345].
%%CITATION = HEP-PH 0307345;%%
%
%\cite{Sugiyama:2003zk}
\bibitem{Sugiyama:2003zk}
J.~Sugiyama, T.~Doi and M.~Oka,
%``Penta-quark baryon from the QCD sum rule,''
Phys.\ Lett.\ B {\bf 581}, 167 (2004)
[arXiv:hep-ph/0309271].
%%CITATION = HEP-PH 0309271;%%
%
%\cite{Sasaki:2003gi}
\bibitem{Sasaki:2003gi}
S.~Sasaki,
%``Lattice study of exotic S = +1 baryon,''
Phys.\ Rev.\ Lett.\  {\bf 93}, 152001 (2004) 
[arXiv:hep-lat/0310014].
%%CITATION = HEP-LAT 0310014;%%
%%CITATION = HEP-LAT 0310014;%%
%
%\cite{Lyubovitskij:2001nm}
\bibitem{Lyubovitskij:2001nm}
V.~E.~Lyubovitskij, T.~Gutsche and A.~Faessler,
%``Electromagnetic structure of the nucleon in the perturbative 
%chiral  quark model,''
Phys.\ Rev.\ C {\bf 64}, 065203 (2001)
[arXiv:hep-ph/0105043].
%%CITATION = HEP-PH 0105043;%%
%
%\cite{Lyubovitskij:2000sf}
\bibitem{Lyubovitskij:2000sf}
V.~E.~Lyubovitskij, T.~Gutsche, A.~Faessler and E.~G.~Drukarev,
%``Sigma-term physics in the perturbative chiral quark model,''
Phys.\ Rev.\ D {\bf 63}, 054026 (2001)
[arXiv:hep-ph/0009341]; 
%%CITATION = HEP-PH 0009341;%%
%
%%\cite{Lyubovitskij:2001fv}
%\bibitem{Lyubovitskij:2001fv}
V.~E.~Lyubovitskij, T.~Gutsche, A.~Faessler and R.~Vinh Mau,
%``pi N scattering and electromagnetic corrections 
%in the perturbative  chiral quark model,''
Phys.\ Lett.\ B {\bf 520}, 204 (2001)
[arXiv:hep-ph/0108134]; 
%%CITATION = HEP-PH 0108134;%% 
%
%%\cite{Lyubovitskij:2001zn}
%\bibitem{Lyubovitskij:2001zn}
%V.~E.~Lyubovitskij, T.~Gutsche, A.~Faessler and R.~Vinh Mau,
%``Electromagnetic couplings of the ChPT Lagrangian from 
%the perturbative  chiral quark model,''
Phys.\ Rev.\ C {\bf 65}, 025202 (2002)
[arXiv:hep-ph/0109213]; 
%%CITATION = HEP-PH 0109213;%%
%
%%\cite{Lyubovitskij:2002ng}
%\bibitem{Lyubovitskij:2002ng}
V.~E.~Lyubovitskij, P.~Wang, T.~Gutsche and A.~Faessler,
%``Strange nucleon form factors in the perturbative chiral 
%quark model,''
Phys.\ Rev.\ C {\bf 66}, 055204 (2002)
[arXiv:hep-ph/0207225]; 
%
%%\cite{Simkovic:2001fy}
%\bibitem{Simkovic:2001fy}
F.~Simkovic, V.~E.~Lyubovitskij, T.~Gutsche, A.~Faessler 
and S.~Kovalenko,
%``Neutrino mediated muon electron conversion in nuclei revisited,''
Phys.\ Lett.\ B {\bf 544}, 121 (2002)
[arXiv:hep-ph/0112277];  
%%CITATION = HEP-PH 0112277;%%
%
%%\cite{Pumsa-ard:2003yh}
%\bibitem{Pumsa-ard:2003yh}
K.~Pumsa-ard, V.~E.~Lyubovitskij, T.~Gutsche, A.~Faessler 
and S.~Cheedket,
%``Electromagnetic nucleon Delta transition in the 
%perturbative chiral  quark model,''
Phys.\ Rev.\ C {\bf 68}, 015205 (2003)
[arXiv:hep-ph/0304033];
%%CITATION = HEP-PH 0304033;%%
%
%%\cite{Inoue:2003bk}
%\bibitem{Inoue:2003bk}
T.~Inoue, V.~E.~Lyubovitskij, T.~Gutsche and A.~Faessler,
 %``Updated analysis of meson nucleon sigma terms in the perturbative chiral
%quark model,''
Phys.\ Rev.\ C {\bf 69}, 035207 (2004)
[arXiv:hep-ph/0311275];
%%CITATION = HEP-PH 0311275;%%
%
%%\cite{Cheedket:2002ik}
%\bibitem{Cheedket:2002ik}
S.~Cheedket, V.~E.~Lyubovitskij, T.~Gutsche, A.~Faessler, 
K.~Pumsa-ard and Y.~Yan,
%``Electromagnetic form factors of the baryon octet 
%in the perturbative chiral quark model,''
Eur.\ Phys.\ J.\ A {\bf 20}, 317 (2004) 
[arXiv:hep-ph/0212347]; 
%%CITATION = HEP-PH 0212347;%%
% 
%%\cite{Khosonthongkee:2004qm}
%\bibitem{Khosonthongkee:2004qm}
K.~Khosonthongkee, V.~E.~Lyubovitskij, T.~Gutsche, A.~Faessler, 
K.~Pumsa-ard, S.~Cheedket and Y.~Yan,
%``Axial form factor of the nucleon in the 
%perturbative chiral quark model,''
J.\ Phys.\ G {\bf 30}, 793 (2004)
[arXiv:hep-ph/0403119].
%%CITATION = HEP-PH 0403119;%%
%
%\cite{Inoue:2004}
\bibitem{Inoue:2004}
T.~Inoue, V.~E.~Lyubovitskij, T.~Gutsche and A.~Faessler,
%``Grond-state baryon masses in the perturbative chiral quark model,''
[arXiv:hep-ph/0404051].
%
%\cite{Leutwyler:1980ma}
\bibitem{Leutwyler:1980ma}
H.~Leutwyler,
%``Constant Gauge Fields And Their Quantum Fluctuations,''
Nucl.\ Phys.\ B {\bf 179}, 129 (1981); 
%%CITATION = NUPHA,B179,129;%%
%
%%\cite{Diakonov:1983hh}
%\bibitem{Diakonov:1983hh}
D.~Diakonov and V.~Y.~Petrov,
%``Instanton Based Vacuum From Feynman Variational Principle,''
Nucl.\ Phys.\ B {\bf 245}, 259 (1984); 
%%CITATION = NUPHA,B245,259;%%
%
%%\cite{Efimov:1993ei}
%\bibitem{Efimov:1993ei}
G.~V.~Efimov and M.~A.~Ivanov, 
{\it The Quark Confinement Model of Hadrons}, 
(IOP Publishing, Bristol $\&$ Philadelphia, 1993). 
%
%\cite{Gasser:1987rb}
\bibitem{Gasser:1987rb}
J.~Gasser, M.~E.~Sainio and A.~Svarc,
%``Nucleons With Chiral Loops,''
Nucl.\ Phys.\ B {\bf 307}, 779 (1988).
%%CITATION = NUPHA,B307,779;%%
%
%\cite{Gasser:1982ap}
\bibitem{Gasser:1982ap}
J.~Gasser and H.~Leutwyler,
%``Quark Masses,''
Phys.\ Rept.\  {\bf 87}, 77 (1982).
%%CITATION = PRPLC,87,77;%%
%
%\cite{Maris:1999nt}
\bibitem{Maris:1999nt}
P.~Maris and P.~C.~Tandy,
%``Bethe-Salpeter study of 
%vector meson masses and decay constants,''
Phys.\ Rev.\ C {\bf 60}, 055214 (1999)
[arXiv:nucl-th/9905056]; 
%%CITATION = NUCL-TH 9905056;%
%
%%\cite{Roberts:2000aa}
%\bibitem{Roberts:2000aa}
C.~D.~Roberts and S.~M.~Schmidt,
%``Dyson-Schwinger equations: 
%Density, temperature and continuum strong  QCD,''
Prog.\ Part.\ Nucl.\ Phys.\  {\bf 45}, S1 (2000)
[arXiv:nucl-th/0005064].
%%CITATION = NUCL-TH 0005064;%%},
%
%\cite{Wybourne:2003if}
\bibitem{Wybourne:2003if}
B.~G.~Wybourne,
%``Group theory and the pentaquark,''
arXiv:hep-ph/0307170.
%%CITATION = HEP-PH 0307170;%%
%
%\cite{Close:2003tv}
\bibitem{Close:2003tv}
F.~E.~Close,
%``The end of the constituent quark model?,''
AIP Conf.\ Proc.\  {\bf 717}, 919 (2004)
[arXiv:hep-ph/0311087].
%%CITATION = HEP-PH 0311087;%%
%\cite{Cahn:2003wq}
\bibitem{Cahn:2003wq}
R.~N.~Cahn and G.~H.~Trilling,
%``Experimental limits on the width of the reported Theta(1540)+,'' 
Phys.\ Rev.\ D {\bf 69} 011501 (2004)   
[arXiv:hep-ph/0311245]. 
%%CITATION = HEP-PH 0311245;%%
%\cite{Eidemuller:2005jm}
\bibitem{Eidemuller:2005jm}
  M.~Eidemuller, F.~S.~Navarra, M.~Nielsen and R.~R.~da Silva,
  %``Pentaquark decay width in QCD sum rules,''
  arXiv:hep-ph/0503193.
  %%CITATION = HEP-PH 0503193;%%
%\cite{Diakonov:2005ib}
\bibitem{Diakonov:2005ib}
  D.~Diakonov and V.~Petrov,
  %``Estimate of the Theta+ width in the Relativistic Mean Field
  %Approximation,''
  arXiv:hep-ph/0505201.
  %%CITATION = HEP-PH 0505201;%%
%\cite{Wang:2005ms}
\bibitem{Wang:2005ms}
  Z.~G.~Wang, W.~M.~Yang and S.~L.~Wan,
  %``Decay width of the pentaquark state Theta(1540)+ with QCD sum rules,''
  arXiv:hep-ph/0504151.
  %%CITATION = HEP-PH 0504151;%%
\end{thebibliography}
\end{document}